\documentclass[10pt,final,doublecolumn]{IEEEtran}
\IEEEoverridecommandlockouts
\usepackage{amsmath}
\usepackage{amssymb}
\usepackage{amsthm}
\usepackage{mathrsfs}
\usepackage{latexsym}
\usepackage{graphicx}
\usepackage{bbding}
\usepackage{indentfirst}
\usepackage{cases}
\usepackage{supertabular}
\usepackage{algorithm,algorithmic}
\usepackage{subeqnarray}
\usepackage{color}
\usepackage{bm}
\usepackage{stfloats}

\usepackage{subfigure}
\usepackage{array}
\usepackage{stfloats}
\usepackage{algorithmic}
\usepackage{float}
\usepackage{algorithm}
\allowdisplaybreaks[4]

\begin{document}
\title{Robust Beamforming Design for Integrated Satellite-Terrestrial Maritime Communications in the Presence of Wave Fluctuation}
\author{\IEEEauthorblockN{
\normalsize Kaiwei Xiong, Xiaoming Chen, and Ming Ying}
\thanks{Kaiwei Xiong, Xiaoming Chen, and Ming Ying are with the College of Information Science and Electronic Engineering, Zhejiang University, Hangzhou 310027, China (e-mail:\{xiong\_kaiwei, chen\_xiaoming, and ming\_ying\}@zju.edu.cn).}
}\maketitle

\begin{abstract}
In order to provide wireless services for wide sea area, this paper designs an integrated satellite-terrestrial maritime communication framework. Specifically, the terrestrial base station (TBS) serves near-shore users, while the low earth orbit (LEO) satellite communicates with off-shore users. We aim to improve the overall performance of integrated satellite-terrestrial maritime communication system. Thus, it makes sense to jointly optimize transmit beamforming at the TBS and LEO satellite. Due to sea wave fluctuation, the obtained channel state information (CSI) is often imperfect. In this context, a robust beamforming design algorithm is proposed with the goal of minimizing the total power consumption of integrated satellite-terrestrial maritime communication system while satisfying quality of service (QoS) requirements. Both theoretical analysis and simulation results confirm the effectiveness of proposed algorithm in maritime communications.
	
\end{abstract}
\providecommand{\keywords}[1]{\textbf{\textit{Index terms---}} #1}

\section{Introduction}
Currently, oceanic activities such as offshore oil exploration, maritime environmental monitoring, maritime fisheries, and maritime scientific research are on the rise \cite{Maritime transport1}, \cite{Maritime transport2}. These activities drive a growing demand for maritime communications with reliable high-speed data rates and wide wireless coverage \cite{growing demand1}-\cite{growing demand3}. For example, operational data and navigational information are necessary for the majority of ships to navigate safely, and multimedia communication services are essential for aboard personnel, passengers, and fisherman. Similarly, offshore drilling platforms need to communicate operational data in real time, and buoys are responsible for uploading significant amounts of meteorological data. Moreover, in maritime rescue operations, real-time video communication is often necessary to ensure effective shore-to-vessel and vessel-to-vessel coordination. Therefore, building the maritime communication network for the maritime Internet-of-Things (IoT) \cite{IoT} is of great significance for maritime transportation, production safety and emergency rescue. In recent years, terrestrial communication technology has witnessed remarkable advancements, offering unprecedented data rates and a wide array of new applications. The upcoming sixth generation (6G) era is expected to further enhance capacity by dozens of times \cite{6G1}, \cite{6G2}. Unfortunately, the maritime domain has lagged behind in this communications revolution. Maritime communication still significantly lags behind terrestrial communication.

Mobile terminals at sea primarily rely on maritime LEO satellites \cite{satellite1}, \cite{satellite2} or TBSs for connectivity. Hence, maritime communication systems can be broadly categorized into two groups. The first category encompasses TBS-based communication systems utilizing the ultra high frequency band and the very high frequency band to cover near-shore areas. The second category involves the maritime satellite communication systems operating in the S frequency band and L frequency (VHF) band. Narrowband satellites primarily provide some low communication rate services. However, there is a growing need for satellite broadband communication services. Thus, numerous companies have introduced some satellites with high throughput such as Starlink project and the Jupiter-2. To expand the coverage of maritime communication systems, shore-based and island-based TBSs \cite{TBS} are also employed alongside maritime satellites. There are some existing shore-based systems which can offer some services with low speed data, such as the Automatic Identification System (AIS) and Navigation Telex (NAVTEX) system \cite{TBS service}. However, efforts have still been made by other companies to enhance communication data rates. These include conducting tests on long distance transmission from shore to vessel by using long term evolution (LTE) or global interoperability of microwave access (WiMAX) networks \cite{high rate}. However, the coverage provided by shore-based communication systems is often limited, especially in the remote sea. The dynamic nature of ships and maritime activities necessitates communication systems capable of real-time communication with mobile vessels.

In maritime communications, well-equipped shore-based public communications network offers the convenience of high-speed data services to maritime users onboard ships. However, coverage is still a significant challenge to overcome. In order to meet wide coverage, satellites have become a critical consideration in the next generation of communication network. Through relay cooperation, signal retransmission can be carried out for users who receive poor satellite signals. Recent work has explored the integration of satellite and ocean networks. In \cite{7}, stations based on island and shore are seen as extensions of terrestrial networks to fill the gap in satellite transmission at sea. In \cite{8}, the author studies cooperative transmission in satellite-ground integrated networks and proposes the comprehensive framework for cooperative transmission in satellite-ground systems in the future. While many LEO satellites with high throughput can provide communication services of high speed data and cover remote areas, satellite system resources remain insufficient to meet the rapidly growing maritime service demands. In \cite{Hybrid}, the authors analyze potential application of integrated satellite-ground networks in maritime communications, and point out that existing satellite systems have limited resources and cannot independently meet the increasing bandwidth and reliability requirements of maritime services. Additionally, the high cost of satellite communication is also an important problem that cannot be ignored. Therefore, an efficient integrated satellite-terrestrial maritime communication network is urgently needed. The integrated system can offer a practical solution for maritime networks, which combines the wide coverage advantages of satellites with the high capacity of shore-based systems. The proposed integrated satellite-terrestrial maritime communication network can transfer part of the traffic to the shore-based system, which reduces the demand for satellite system resources, especially decreases the required number of satellites. Therefore, the reduced number of satellites can definitely decrease the overall deployment and operating costs of the entire maritime communication system. Moreover, by combining the advantages of TBS and LEO satellite, the integrated network can improve the performance of the whole system and increase the resource utilization efficiency.

Energy efficiency plays a crucial role in maritime communication because of the substantial loss experienced in maritime paths and the limited availability of TBS. Maritime communication networks must cover a wide area and often require the high-power TBS for remote transmission, which leads to less efficient energy use. Hence, the design of efficient and energy-saving allocation strategies holds great significance. At present, the resource allocation technology of terrestrial network has been widely studied. In \cite{10}, a cooperative power allocation and user scheduling scheme is proposed for hybrid delay services, employing a cross-layer approach. Reference \cite{9} introduces a dynamic programming-based algorithm that combines power allocation and user scheduling for multi-user multiple-input multiple-output (MIMO) systems. More recently, the authors in \cite{11} present a pilot and scheduling allocation scheme for massive MIMO systems, aiming to maximize users capacity while ensuring quality of service (QoS) requirements. Besides, in \cite{12}, energy efficiency is improved by using specific characteristics of sea lanes and user mobility. In particular, it is possible to estimate their large-scale CSI by utilizing the channel information of maritime users.

However, the assumption behind these research is perfect CSI, which is unfeasible in actual maritime communication. In general, CSI acquisition has channel phase error due to the error and delay when the device conveys CSI to the satellite through gateway. Radio communication at sea is subject to a variety of interference, including signal scattering caused by rain, antenna vibration caused by waves, and interference of mutual reflection. Thus, it becomes necessary to take into account the influence of imperfect CSI. In fact, several researches have been conducted to specifically investigate multi-beam satellite communication in the presence of uncertain channel conditions. The work \cite{13} is devoted to solving the beamforming problem in terrestrial-satellite integrated networks. By adopting beamforming technology and considering the influence of channel uncertainty and interference, the performance and reliability of the system are improved. In \cite{14}, it is proposed a scheme for robust beamforming design to meet the requirements of multi-group multicast communication and maintain the reliability of the system under the interruption limitation. To maximize the total secrecy of the system, robust beamforming schemes are proposed in \cite{15} and \cite{16}, based on the assumption of the additive norm bounded estimation error and the non-orthogonal multi-access multi-beam access technique in satellite communication. Furthermore, the work \cite{17} is dedicated to optimizing the resource utilization efficiency of satellite communication systems. By introducing a robust beamforming technology and the reasonable resource allocation strategy, the performance and reliability of the system are improved. Although the above works have modeled the CSI error, these CSI error models are not constructed for specific maritime scenarios and do not take into account the characteristics of scenarios. Our work is specially designed for maritime communications, which focuses on the characteristics of the ocean, such as wave fluctuations. Therefore, the proposed CSI error model is based on the characteristics of the ocean and is designed specially for the maritime communication, which is applied to our proposed integrated satellite-terrestrial maritime communication system. The integrated satellite-terrestrial maritime communication network offers extensive coverage, stable and reliable communication capabilities, and establishes a viable foundational framework for the implementing of maritime IoT network framework. Furthermore, the integrated satellite-terrestrial maritime communication network provides connectivity and communication capabilities for the maritime IoT. In order to provide extensive and seamless wireless network coverage and high-speed communication services in the maritime environment, it is desired to increase the transmit power, which also leads to severe co-channel interference. Considering limited power resource, we advocate to jointly design the beamforming at TBS and LEO satellite to minimize the transmit power of the system.

This paper focuses on expanding the coverage of maritime communication network and ensure communication quality for users. We propose an integrated satellite-terrestrial maritime communication system, where TBS and LEO satellites cooperatively provide wide-area wireless services. In addition, we consider imperfect CSI of the maritime channels caused by wave fluctuation. To reduce interference, and ensure the communication quality of users, we investigate a robust beamforming design that combines LEO satellites and TBS. The aim is to minimize the total transmit power while guaranteeing the minimum data rate requirements for maritime users. The main contributions of this paper can be summarized as following aspects.

\begin{enumerate}
	
	\item We propose an integrated satellite-terrestrial maritime communication system, which combines a TBS with multiple antennas to provide network coverage for near-shore users, and a maritime LEO satellite to provide network coverage for off-shore users.
	
	\item We analyze the impacts of sea wave fluctuation on CSI acquisition, and accurately model the CSI errors at the LEO satellite and TBS.
	
	\item We propose a robust beamforming design algorithm for integrated satellite-terrestrial maritime communications based on the obtained CSI. Both theoretical analysis and simulation results confirm the effectiveness of the proposed algorithm.
	
\end{enumerate}

The rest of this article is organized as follows. Section \uppercase\expandafter{\romannumeral2} introduces an integrated satellite-terrestrial maritime communication system. In section \uppercase\expandafter{\romannumeral3}, the transmit beamforming at the TBS and LEO satellite is jointly designed to minimize the overall power consumption while satisfying QoS requirements. Section \uppercase\expandafter{\romannumeral4} provides simulation results to testify the effectiveness of proposed algorithm. Finally, section \uppercase\expandafter{\romannumeral5} concludes this paper.

\emph{Notations}: This article represents matrices and vectors by bold letters, $\|\cdot\|$, $(\cdot)^H$, $|\cdot|$, $\|\cdot\|_F$ to denote $L_2$-norm of a vector, conjugate transpose, absolute value and the Frobenius norm of a matrix, respectively. $J_0(\cdot)$, $J_1(\cdot)$, and $J_3(\cdot)$ are employed to represent the zeroth, first and third order of the first-kind of Bessel function, respectively. $\text{Rank}(\cdot)$ and $\text{tr}(\cdot)$ denote the rank and trace of a matrix, and $\odot$ denotes  the Hadamard product. ${{\mathbb{C}}^{m\times n}}$ denotes the set of $m\times n$ dimensional complex matrices, $[\cdot]_{n,m}$ denotes the $[n,m]$th element of a matrix, $\mathcal{CN}(\mu,\sigma^2)$ and $\mathcal{N}(\mu,\sigma^2)$ are denote the complex Gaussian distribution and the Gaussian distribution with mean $\mu$ and variance $\sigma^2$, respectively.

\section{System Model}

\subsection{System Description}
\begin{figure}
		\centering
		\includegraphics [width=0.5\textwidth] {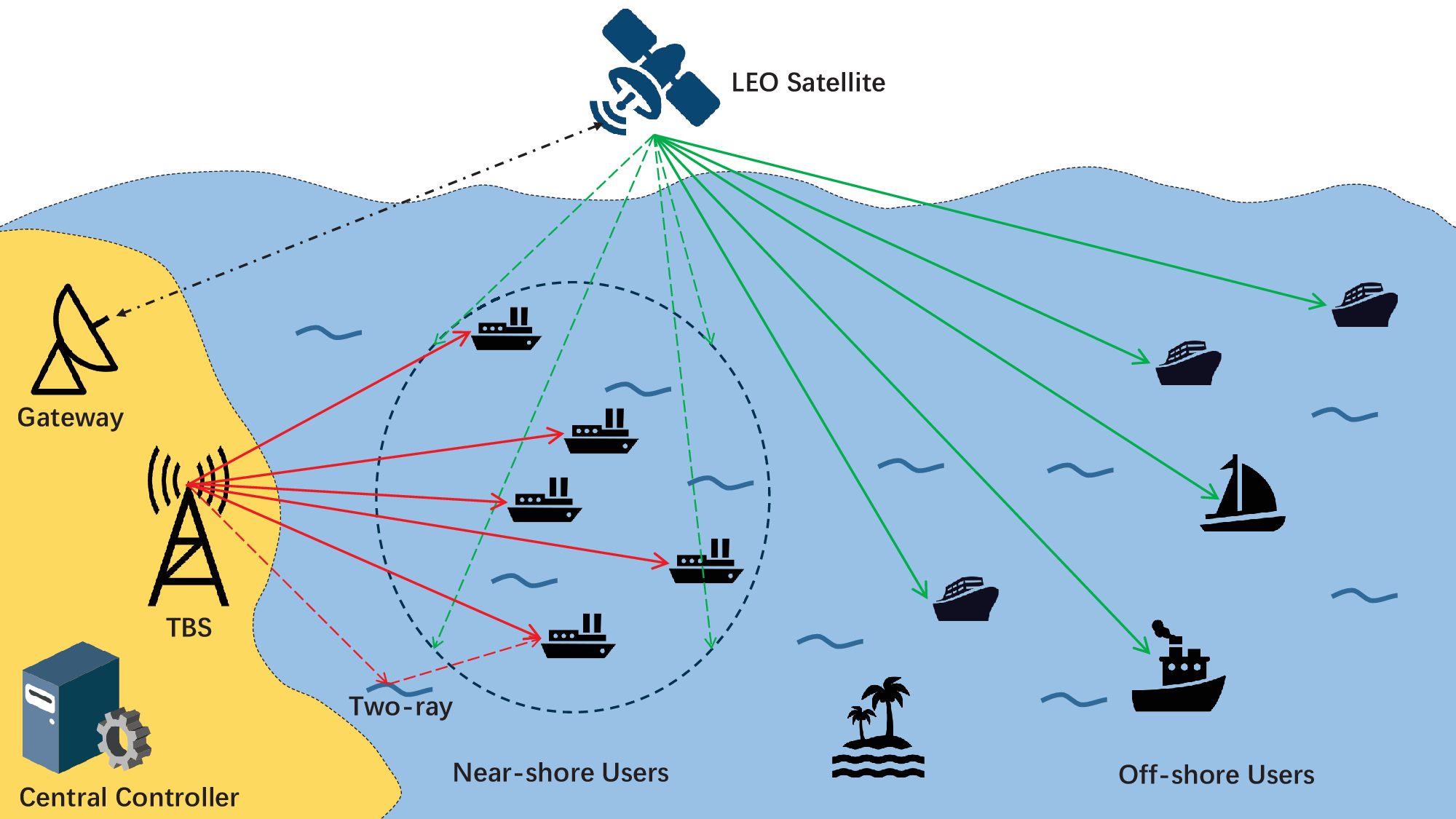}
		\caption {Illustration of an integrated satellite-terrestrial communication system for maritime purposes, where the TBS and LEO satellite provide wide-area services in a coordinated manner.}
		\label{system}
\end{figure}

Consider an integrated satellite-terrestrial maritime communication system shown in Fig. $\ref{system}$, where a LEO satellite and a shore-based TBS cooperatively provide communication services to various maritime users. In general, maritime communication systems work in very high frequency (VHF) band from 30 to 300 MHz. Thus, we consider that the LEO satellite and TBS work over the same spectrum in the VHF band \cite{survey}. Considering the obstruction of sea surface due to the curvature of the earth, we use the trigonometric function to calculate the maximum line of sight (LoS) transmission distance between TBS and maritime user as follows

\begin{equation}
	L_{\max}=\sqrt{H_t^{2}+2H_tR_e}+\sqrt{H_r^{2}+2H_rR_e},
\end{equation}
where $H_t$ is the height of TBS, $H_r$ denotes the effective antennas height of user, and $R_e$ is the Earth radius. For clarity, we define the near-shore users as those whose distance to the TBS is less than $L_{\max}$ and off-shore users as those whose distance to the TBS is more than $L_{\max}$. The near-shore users communicate with the TBS, while the off-shore users communicate with the maritime LEO satellite. The TBS on the shore equipped with $M_1$ antennas transmits the desired signals to $K_1$ near-shore users, and the LEO satellite equipped with $M_2$ antennas serves $K_2$ off-shore users. For simplicity, we assume that all the maritime terminals in the system have a single antenna each.

\subsection{Channel Model}
In general, maritime wireless channel is location independent due to the lack of obstacles and the sparsity of scatters. Thus, maritime channel is mainly composed of reflected path on the ocean surface and LoS path, namely two-ray channel model \cite{two ray}. As the scattering effects are significantly weaker compared to the first two main paths in the two-ray model, we ignore the impacts resulted by the diffusion paths of the rough ocean surface. In general, the communication distance will affect the transmit power, and the maritime path loss also affects the energy efficiency directly. The channel gain from the TBS to near-shore user $i$ can be modeled as

\begin{equation}\label{channel1}
	\begin{aligned}
		\mathbf{h}_{1,i}=\bigg[\frac{\lambda }{2\pi d_i}\sin(\frac{2\pi H_tH_u}{\lambda d_i})\bigg]\mathbf{a}(\theta_1 ),
	\end{aligned}
\end{equation}
where $d_i$ is the distance between TBS and the user $i$, $H_u$ is the effective antenna height of users, and $\mathbf{a}(\theta_1)$ denotes the antenna array response of the link from TBS to the user $i$, which can be expressed as

\begin{equation}\label{antenna array response}
	\begin{aligned}
		\mathbf{a}(\theta_1)=[1,e^{-j(\frac{2\pi b}{\lambda}\cos\theta_1)},\dots, e^{-j(\frac{2\pi b(M_1-1)}{\lambda}\cos\theta_1)}],
	\end{aligned}
\end{equation}
where $\theta_1$ denotes the angle of departure, and $b$ is the distance between two adjacent antennas at the TBS. Since the distance between the transmitter and receiver is large enough, we assume that the departure angles of all links are the same. On the other hand, considering the electromagnetic propagation characteristics from the LEO satellite to the off-shore users \cite{two ray}-\cite{ym}, the space-to-sea channel associated with off-shore user $m$ can be modeled as

\begin{equation}\label{channel2}
	\begin{aligned}
		\mathbf{h}_{2,m}=g_m\sqrt{P_{m}}\bigg(\sqrt{\frac{1}{1+K}}\mathbf{h}_{2,m}^{\text{NLoS}}+\sqrt{\frac{K}{1+K}}\mathbf{h}_{2,m}^{\text{LoS}}\bigg),
	\end{aligned}
\end{equation}
where $g_m$ is the large-scale fading factor, which is given by

\begin{equation}\label{channel2}
	\begin{aligned}
		g_m=\mathbf{r}_m\sqrt{r_{\text{FSL}}G_{m}},
	\end{aligned}
\end{equation}
where $G_m$ is the receive antenna gain of off-shore user $m$, and $r_{\text{FSL}}$ denotes the free space loss (FSL) as follows

\begin{equation}\label{FSL}
	\begin{aligned}
		\ r_{\text{FSL}}=\frac{\lambda^{2}}{(4\pi )^{2}D_m},
	\end{aligned}
\end{equation}
where $D_m$ is the distance between LEO satellite and the user $m$. Furthermore, $\mathbf{r}_m$ is the rain attenuation coefficient vector. In the form of dB, the rain attenuation model $r_{i}^{\text{dB}}(k)=20\log_{10}r_{i}(k)$ follows a log-normal distribution with $\ln(r_{i}^{\text{dB}})\sim\mathcal{C}\mathcal{N}(\mu_r, \sigma_r^2)$ \cite{rain}, where $\sigma_r^2$ and $\mu_r$ denote the scale parameter and log-normal location, respectively. Moreover, $K$ denotes the Rician fading factor of the space-to-sea channel, $\mathbf{h}_{2,m}^{\text{NLoS}}$ is the non line of sight (NLoS) component of $\mathbf{h}_{2,m}$, which is an identically and independently distributed variable following distribution $\mathcal{C}\mathcal{N}(0,1)$, and $\mathbf{h}_{2,m}^{\text{LoS}}$ is the LoS component of $\mathbf{h}_{2,m}$, which is given by

\begin{equation}\label{h_L}
	\begin{aligned}
		\mathbf{h}_{2,m}^{\text{LoS}}=[1,e^{-j(\frac{2\pi b}{\lambda}\cos\theta_2)},\dots,
		e^{-j(\frac{2\pi b(M_2-1)}{\lambda}\cos\theta_2)}],
	\end{aligned}
\end{equation}
where $\theta_2$ is the angle of departure of the link from the LEO satellite to the off-shore user $m$. Besides, the LEO satellite transmit antenna gain $P_{m}$ can be expressed as \cite{rain}

\begin{equation}\label{P_m}
	\begin{aligned}
		P_{m}=G_{\max}\bigg(\frac{J_1(\psi_m)}{2\psi_m}+36\frac{J_3(\psi_m)}{\psi_{m}^{3}}\bigg)^{2},
	\end{aligned}
\end{equation}
where $G_{\max}$ is the maximal LEO satellite antenna gain, $J_3(\cdot)$ and $J_1(\cdot)$ denote the third and first orders of the first-kind of Bessel functions, respectively, and $\psi_m=2.07123\sin{\varphi_{\text{sm}}}/\sin{\varphi_{\text{3dB},m}}$ with $\varphi_{\text{sm}}$ being the angle of the satellite beam boresight to the user $m$, and $\varphi_{\text{3dB},m}$ being the 3dB angle.

\subsection{Signal Model}
In order to improve the overall performance of maritime communications, the TBS and the LEO satellite coordinate their transmit signals via spatial beamforming. Specifically, the transmit signals $\mathbf{x}_1$ for near-shore users and $\mathbf{x}_2$ for off-shore users can be expressed as

\begin{equation}\label{signal1}
	\begin{aligned}
		\mathbf{x}_1=\sum_{i=1}^{K_1}\mathbf{w}_is_{1,i},
	\end{aligned}
\end{equation}

\begin{equation}\label{signal2}
	\begin{aligned}
		\mathbf{x}_2=\sum_{m=1}^{K_2}\mathbf{v}_ms_{2,m},
	\end{aligned}
\end{equation}
where $\mathbf{w}_i$ and $\mathbf{v}_m$ are the designed beams for the near-shore user $i$ and the off-shore user $m$, respectively. The $s_{1,i}$ and $s_{2,m}$ denote the signals for the near-shore user $i$ and the off-shore user $m$ with unit norm, respectively. For the near-shore users, they receive the desired signals from TBS, and also suffer the interference from the LEO satellite due to wide space-to-sea beam. Thus, the received signal at the near-shore user $i$ is formulated as

\begin{equation}\label{y_i}
	\begin{aligned}
		y_{1,i}&=\mathbf{h}_{1,i}\mathbf{x}_1+\hat{\mathbf{h}}_{2,i}\mathbf{x}_2+n_{1,i}  \\
		   &=\underbrace{\mathbf{h}_{1,i}\mathbf{w}_i s_{1,i}}_{\text{Desired  Signal}}+\underbrace{\sum_{j=1,j\neq i}^{K_1} \mathbf{h}_{1,i}\mathbf{w}_js_{1,j}}_{\text{Inter-device Interference}}+ \\
		   &\underbrace{\sum_{m=1}^{K_2} \hat{\mathbf{h}}_{2,i}\mathbf{v}_ms_{2,m}}_{\text{Intra-device Interference}}+\underbrace{n_{1,i}}_{\text{AWGN}},
	\end{aligned}
\end{equation}
where $\hat{\mathbf{h}}_{2,i}$ denotes the space-to-sea channel gain from LEO satellite to the near-shore user $i$, and $n_{1,i}$ is the additive white Gaussian noise (AWGN) at the near-shore user $i$ with variance $\sigma_{1,i}^{2}$. Hence the signal to interference plus noise ratio (SINR) at the near-shore user $i$ can be computed as

\begin{equation}\label{SINR_i}
	\begin{aligned}
	\text{SINR}_{1,i}=\frac{|\mathbf{h}_{1,i}\mathbf{w}_i|^2}{\sum_{j=1,j\ne   i}^{K_1}|\mathbf{h}_{1,i}\mathbf{w}_j|^2+\sum_{m=1}^{K_2}|\hat{\mathbf{h}}_{2,i}\mathbf{v}_m|^2+\sigma_{1,i}^{2}}{\color{blue}.}
	\end{aligned}
\end{equation}
Accordingly, the achievable date rate of near-shore user $i$ can be expressed as

\begin{equation}\label{R_i}
	\begin{aligned}
		R_{1,i}=\log_2(1+\text{SINR}_{1,i}){\color{blue}.}
	\end{aligned}
\end{equation}
Due to the curvature of the earth, the off-shore users can not receive the signals from the TBS, and only from the LEO satellite. Thus, the received signal at the off-shore user $m$ is formulated as

\begin{equation}\label{y_m}
	\begin{aligned}
		y_{2,m}&=\mathbf{h}_{2,m}\mathbf{x}_2+n_{2,m} \\
		&=\underbrace{\mathbf{h}_{2,m}\mathbf{v}_m s_{2,m}}_{\text{Desired Signal}}+\underbrace{ \sum_{k=1,k\neq m}^{K_2}\mathbf{h}_{2,m}\mathbf{v}_ks_{2,k}}_{\text{Inter-device Interference}}+\underbrace{n_{2,m}}_{\text{AWGN}},
	\end{aligned}
\end{equation}
where $n_{2,m}$ is the AWGN at the off-shore user $m$ with variance $\sigma_{2,m}^{2}$. Then the SINR at the off-shore user $m$ can be computed as

\begin{equation}\label{SINR_m}
	\begin{aligned}
		\text{SINR}_{2,m}=\frac{|\mathbf{h}_{2,m}\mathbf{v}_m|^2}{\sum_{k=1,k\ne   m}^{K_2}|\mathbf{h}_{2,m}\mathbf{v}_k|^2+\sigma_{2,m}^{2}}{\color{blue}.}
	\end{aligned}
\end{equation}
Similarly, the achievable date rate of off-shore user $m$ can be expressed as

\begin{equation}\label{R_m}
	\begin{aligned}
		R_{2,m}=\log_2(1+\text{SINR}_{2,m}){\color{blue}.}
	\end{aligned}
\end{equation}
It is not difficult to find that the beam $\mathbf{w}_i$ and $\mathbf{v}_m$ have a great impact on the overall performance of the integrated satellite-terrestrial maritime communication system. In the maritime communication system, the power of LEO satellite and TBS is often limited. Thus, we aim to design the beamforming in the next section to minimize the total power at the transmitter while ensuring the communication quality of maritime users.

\section{Robust Beamforming Design}
In this section, we first model the acquisition of CSI in the context of maritime communication, and then design transmit beams for TBS and LEO satellite based on the obtained CSI.

\subsection{CSI Acquisition}
In general, the maritime electromagnetic propagation is influenced by a variety of time varying factors such as atmospheric conditions and sea surface. The height of sea surface is affected by tidal movement and wave fluctuation. Wherein, the wave fluctuation causes the change of azimuth angle of maritime wireless channel. As a result, there exists angle mismatch between the actual CSI and the obtained CSI, which is given by

\begin{equation}\label{mismatch}
	\begin{aligned}
		\theta_1=\bar{\theta}_1+\Delta \theta_1, \\
		\theta_2=\bar{\theta}_2+\Delta \theta_2,
	\end{aligned}
\end{equation}
where $\theta_i$ is the actual angle of departure, $\bar{\theta}_i$ is the obtained angle of departure, and $\Delta \theta_i$ is the error of angle of departure, which is constrained within a hyper-spherical region centered at the origin with radii of $\varepsilon_i$. Mathematically, the error of angle of departure are bounded as

\begin{equation}
	\begin{aligned}
		\parallel \Delta \theta_1 	\parallel_2\leq \varepsilon_1, \\
	    \parallel \Delta \theta_2 	\parallel_2\leq \varepsilon_2.
	\end{aligned}
\end{equation}

We assume that the angle errors are unknown but the corresponding bounds $\varepsilon_1$ and $\varepsilon_2$ are available. Hence, the relationship between actual CSI and obtained CSI can be expressed as

\begin{equation}\label{h1}
	\begin{aligned}
		\mathbf{h}_{1,i}=\bar{\mathbf{h}}_{1,i}+\Delta \mathbf{h}_{1,i},
	\end{aligned}
\end{equation}

\begin{equation}\label{h2}
	\begin{aligned}
		\mathbf{h}_{2,m}=\bar{\mathbf{h}}_{2,m}+\Delta \mathbf{h}_{2,m},
	\end{aligned}
\end{equation}
where $\bar{\mathbf{h}}_{1,i}$ and $\bar{\mathbf{h}}_{2,m}$ denote the obtained value of  $\mathbf{h}_{1,i}$ and $\mathbf{h}_{2,m}$, respectively, and they can be expressed as

\begin{equation}\label{error1}
	\begin{aligned}
		\bar{\mathbf{h}}_{1,i}=\bigg[\frac{\lambda }{2\pi d_i}\sin(\frac{2\pi H_tH_u}{\lambda d_i})\bigg]\bar{\mathbf{a}}(\theta_1 ),
	\end{aligned}
\end{equation}

\begin{equation}\label{error2}
	\begin{aligned}
		\bar{\mathbf{h}}_{2,m}=g_m\sqrt{P_{m}}\bigg(\sqrt{\frac{1}{1+K}}\mathbf{h}_{2,m}^{\text{NLoS}}+\sqrt{\frac{K}{1+K}}\bar{\mathbf{h}}_{2,m}^{\text{LoS}}\bigg).
	\end{aligned}
\end{equation}

To derive the CSI error $\Delta{\mathbf{h}_{1,i}}$, according to the angle mismatch in ($\ref{mismatch}$), we take the first-order Taylor expansion on the $k$-th element of antenna array response in ($\ref{antenna array response}$). Then, we have

\begin{equation}\label{error}
	\begin{aligned}
	&\exp\bigg(-j\frac{2\pi bk}{\lambda }\cos(\bar{\theta}_1+\Delta \theta_1)\bigg) \\
	\approx 	&\underbrace{\exp\bigg(-j\frac{2\pi bk}{\lambda }\cos(\bar{\theta}_1)\bigg)}_{\text{$\bar{\mathbf{A}_k}(\theta_1 )$}}+\\
	&\underbrace{\bigg(j\frac{2\pi bk}{\lambda }\sin(\bar{\theta}_1)\Delta \theta_1\bigg)\times \exp\bigg(-j\frac{2\pi bk}{\lambda }\cos(\bar{\theta}_1)\bigg)}_{\text{$\Delta \mathbf{a}_k(\theta_1 )$}}.
	\end{aligned}
\end{equation}
Further we redefine $\Delta \mathbf{a}(\theta_1 )$ as

\begin{equation}\label{CSI error}
	\begin{aligned}
		\Delta \mathbf{a}(\theta_1 )\triangleq \bar{\mathbf{a}}(\theta_1 ) \odot \sin(\bar{\theta }_1)[1,\dots ,j\frac{2\pi b}{\lambda}(k-1)] \Delta \theta_1{\color{blue}.}
	\end{aligned}
\end{equation}

Taking each element of the antenna array response matrix applied by ($\ref{CSI error}$) and plugging it into channel, we can get the CSI error $\Delta \mathbf{h}_{1,i}$ as follows

\begin{equation}
	\begin{aligned}
		\Delta \mathbf{h}_{1,i}&=\bar{\mathbf{h}}_{1,i} \Delta \mathbf{a}(\theta_1)\\&=\bar{\mathbf{h}}_{1,i} \bar{\mathbf{a}}(\theta_1 ) \odot \sin(\bar{\theta }_1)[1,\dots ,j\frac{2\pi b}{\lambda}(M_1-1)] \Delta \theta_1.
	\end{aligned}
\end{equation}
By introducing a new variable $\mathbf{n}_1=\bar{\mathbf{h}}_{1,i} \bar{\mathbf{a}}(\theta_1 ) \odot \sin(\bar{\theta }_1)[1,\dots ,j\frac{2\pi b}{\lambda}(M_1-1)]$, where $\mathbf{n}_1$ is constrained within $\xi_1$, the CSI error can be bounded as

\begin{equation}
	\begin{aligned}
		\parallel \Delta \mathbf{h}_{1,i} \parallel_2=\parallel \mathbf{n}_1 \Delta \theta_1 \parallel_2 \leq \xi_1\varepsilon_1.
	\end{aligned}
\end{equation}
Similar to $\Delta \mathbf{h}_{1,i}$, $\Delta \mathbf{h}_{2,m}$ can be approximated as

\begin{equation}\label{error1}
	\begin{aligned}
		\Delta \mathbf{h}_{2,m}=g_m\sqrt{P_m} \sqrt{\frac{K}{1+K}}\Delta \mathbf{h}_{2,m}^{\text{LoS}},
	\end{aligned}
\end{equation}
where $\Delta \mathbf{h}_{2,m}^{\text{LoS}}=\bar{\mathbf{h}}_{2,m}^{\text{LoS}} \bar{\mathbf{a}}(\theta_2) \odot \sin(\bar{\theta}_2)[1,\dots ,j\frac{2\pi b}{\lambda}(M_2-1)] \Delta \theta_2$. Also by introducing a new variable $\mathbf{n}_2=\bar{\mathbf{h}}_{2,m}^{\text{LoS}} \bar{\mathbf{a}}(\theta_2) \odot \sin(\bar{\theta }_2)[1,\dots ,j\frac{2\pi b}{\lambda}(k-1)]$, the CSI error can be bounded as
\begin{equation}
	\begin{aligned}
		\parallel\Delta\mathbf{h}_{2,m}\parallel_2= \bigg|\bigg|
		 g_m\sqrt{P_m}\sqrt{\frac{K}{1+K}}\mathbf{a}_2 \Delta \theta_2 \bigg|\bigg|_2 \leq \xi_2\varepsilon_2.
	\end{aligned}
\end{equation}

\subsection{Robust Design}

Due to wave fluctuation, the obtained CSI is usually imperfect in maritime communications. In this context, we attempt to design a robust beamforming scheme to guarantee QoS of near-shore and off-shore users in the worst case. Specifically, the optimization problem can be formulated as

\begin{subequations}
	\begin{eqnarray}
		\mathcal{Q}1:	\underset{\mathbf{w}_i,\mathbf{v}_m}{\mathop{\text{min}}}\,\!\!\!\!\!\!\!\!\!&&\!\!\!\sum_{i=1}^{K_1}\alpha_{1,i}\parallel \mathbf{w}_i \parallel ^2+\sum_{m=1}^{K_2}\alpha_{2,m}\parallel \mathbf{v}_m \parallel ^2 \label{OP2obj}\\
		\textrm{s.t.}&&\!\! \underset{\parallel \Delta \mathbf{h}_{1,i}\parallel\leq\epsilon_1}{\min} R_{1,i}\geq\gamma_1,\label{OP2st1}\\
		&&\!\! \underset{\parallel \Delta \mathbf{h}_{2,m}\parallel\leq\epsilon_2}{\min} R_{2,m}\geq\gamma_2,\label{OP2st2}\\
		&&\!\! \sum_{i=1}^{K_1}\parallel \mathbf{w}_i\parallel^2\leq P_T,\label{OP2st3}\\
		&&\!\! \bigg[\sum_{m=1}^{K_2}\mathbf{v}_m\mathbf{v}_m^H \bigg]_{k,k}\leq P_S,\quad \forall k \label{OP2st4}
	\end{eqnarray}
\end{subequations}
where $\alpha_{1,i}$ and $\alpha_{2,m}$ are the weights of the near-shore users and off-shore users, $\gamma_1$ and $\gamma_2$ are the minimum data transmission rates required by near-shore users and off-shore users, and $P_T$ and $P_S$ are the maximum transmit power of TBS and a single antenna of LEO satellite, respectively. Obviously, the problem ($\mathcal{Q}1$) is non-convex, and it is difficult to get the optimal solution directly. On the one hand, the constraints ($\ref{OP2st1}$) and ($\ref{OP2st2}$) involve an infinite number of inequality constraints, as these constraints must be set up in any possible realizations of $\Delta \mathbf{h}_{1,i}$ and $\Delta \mathbf{h}_{2,m}$. On the other hand, the optimization variables $\mathbf{w}_i$ and $\mathbf{v}_m$ are highly coupled in the expressions of $R_{1,i}$ and $R_{2,m}$, which are difficult to deal with. We take some transformations for the optimization problem to make it more tractable. Firstly, we introduce auxiliary variables $\mathbf{W}_i=\mathbf{w}_i\mathbf{w}_i^H \in \mathbb{C}^{M_1\times M_1}$, and $\mathbf{V}_m=\mathbf{v}_m\mathbf{v}_m^H \in \mathbb{C}^{M_2\times M_2}$, then the problem is equivalent to the following form

\begin{subequations}
	\begin{eqnarray}
		\mathcal{Q}2:	\underset{\mathbf{W}_i,\mathbf{V}_m}{\mathop{\text{min}}}\,\!\!\!\!\!&&\!\!\!\sum_{i=1}^{K_1}\alpha_{1,i}\text{tr}(\mathbf{W}_i)+\sum_{m=1}^{K_2}\alpha_{2,m}\text{tr}(\mathbf{V}_m) \label{OP4obj}\\
		\textrm{s.t.}&&\!\! \underset{\parallel \Delta \mathbf{h}_{1,i}\parallel\leq\epsilon_1}{\min} R_{1,i}\geq\gamma_1,\label{OP4st1}\\
		&&\!\! \underset{\parallel \Delta \mathbf{h}_{2,m}\parallel\leq\epsilon_2}{\min} R_{2,m}\geq\gamma_2,\label{OP4st2}\\
		&&\!\! \sum_{i=1}^{K_1}\parallel \mathbf{w}_i\parallel^2\leq P_T,\label{OP4st3}\\
		&&\!\! \bigg[\sum_{m=1}^{K_2}\mathbf{V}_m \bigg]_{k,k}\leq P_S,\quad \forall k \label{OP4st4}\\
		&&\!\! \mathbf{W}_i\succeq 0,\text{rank}(\mathbf{W}_i)=1,\forall i,\label{OP4st5}\\
		&&\!\! \mathbf{V}_m\succeq 0,\text{rank}(\mathbf{V}_m)=1,\forall m\label{OP4st6}
	\end{eqnarray}
\end{subequations}

\begin{figure*}[hb] 
	\centering
	\rule[-10pt]{18cm}{0.07em}
	\begin{equation}
		\underset{\parallel \Delta \mathbf{h}_{1,i}\parallel\leq\epsilon_1}{\min}\frac{\mathbf{h}_{1,i}^H\mathbf{W}_i\mathbf{h}_{1,i}}{\sum_{j=1,j\ne   i}^{K_1} \mathbf{h}_{1,i}^H\mathbf{W}_j \mathbf{h}_{1,i}+\sum_{m=1}^{K_2}\hat{\mathbf{h}}_{2,i}^H\mathbf{V}_m\hat{\mathbf{h}}_{2,i}+\sigma_{1,i}^{2}}\geq 2^{\gamma_1}-1
		\label{eq_3}
	\end{equation}
\end{figure*}

\begin{figure*}[hb] 
	\centering
	\begin{equation}
		\underset{\parallel \Delta \mathbf{h}_{2,m}\parallel\leq\epsilon_2}{\min}\frac{\mathbf{h}_{2,m}^H\mathbf{V}_m\mathbf{h}_{2,m}}{\sum_{k=1,k\ne   m}^{K_2}\mathbf{h}_{2,m}^H\mathbf{V}_k\mathbf{h}_{2,m}+\sigma_{2,m}^{2}}\geq 2^{\gamma_2}-1
		\label{eq_4}
	\end{equation}
\end{figure*}

Despite the transformation mentioned above, the problem ($\mathcal{Q}2$) remains intractable, primarily due to the inclusion of an limitless number of constraints in ($\ref{OP4st1}$) – ($\ref{OP4st2}$). To convert these two constraints into an equivalent problem with a finite number of constraints, we need to use the following lemma.

\emph{Lemma 1} (S-procedure): We define the following quadratic function as
 \begin{equation}
 	\begin{aligned}
 		f_k(\mathbf{x})=\mathbf{x}^H\mathbf{A}_k\mathbf{x}+2\text{Re}\lbrace\mathbf{B}_k^H\mathbf{x}\rbrace+c_k,k=1,2,\dots,n
 	\end{aligned}
 \end{equation}
where $\mathbf{x}\in \mathbb{C}^{N\times 1}$ is variable, $\mathbf{A}_k=\mathbf{A}_k^H\in\mathbb{C}^{N\times N}$ can be any Hermitian matrix, vector $\mathbf{b}_k\in \mathbb{C}^{N\times 1}$, and scalar $c_k$, $k=1,2,\dots,n$. Only if there exists $\mu_k \ge 0, k=1,2,\dots,n$, we can get the condition $f_i(\mathbf{x})\ge 0 \Rightarrow f_0(\mathbf{x})\ge 0$, such that

\begin{equation}
	\begin{aligned}
			\sum_{k=1}^{n}\mu_k{
				\left[ \begin{array}{ccc}
					\mathbf{A}_k & \mathbf{b}_k \\
					\mathbf{b}_k^H & c_k
				\end{array}
				\right ]}-{
				\left[ \begin{array}{ccc}
					\mathbf{A}_i & \mathbf{b}_i \\
					\mathbf{b}_i^H & c_i
				\end{array}
				\right ]}\succeq 0.
	\end{aligned}
\end{equation}
Note that the current constraints ($\ref{OP4st1}$) and ($\ref{OP4st2}$) cannot be applied directly to the S-procedure, we perform some matrix transformations and equality transformations \cite{15}, \cite{S}. By substituting ($\ref{SINR_i}$) - ($\ref{R_i}$) and ($\ref{SINR_m}$) - ($\ref{R_m}$) into constraints ($\ref{OP4st1}$) and ($\ref{OP4st2}$), the constraints are transformed as ($\ref{eq_3}$) and ($\ref{eq_4}$) at the bottom of the next page. The essence of constraints ($\ref{eq_3}$) and ($\ref{eq_4}$) is that the two inequality always hold when all channel CSI errors are satisfied, i.e., both inequalities hold in all realizable cases at the same time. Hence, the constraint ($\ref{eq_3}$) can be rewritten as

\begin{equation}\label{zhuanhua1}
	\begin{aligned}
		\mathbf{h}_{1,i}^H\mathbf{W}_i\mathbf{h}_{1,i}-(2^\gamma_1-1)\bigg(\sum_{j=1,j\ne   i}^{K_1} \mathbf{h}_{1,i}^H\mathbf{W}_j \mathbf{h}_{1,i}+\\
		\sum_{m=1}^{K_2}\hat{\mathbf{h}}_{2,i}^H\mathbf{V}_m\hat{\mathbf{h}}_{2,i}+\sigma_{1,i}^{2}\bigg) \geq 0,
	\end{aligned}
\end{equation}

\begin{equation}
	\begin{aligned}
		\Delta \mathbf{h}_{1,i}^H\times(-\mathbf{I}_N)\times\Delta \mathbf{h}_{1,i}  \geq 0,\forall \Delta \mathbf{h}_{1,i}.
	\end{aligned}
\end{equation}
Then we combine the like terms and introduce auxiliary variable $\mathbf{N}_1=\mathbf{W}_i-(2^\gamma_1-1)\sum_{j=1,j\ne i}^{K_1}\mathbf{W}_j$. Moreover, by substituting ($\ref{h1}$) to ($\ref{zhuanhua1}$), the constraint ($\ref{eq_3}$) can be transformed as

\begin{equation}
	\begin{aligned}
		\Delta \mathbf{h}_{1,i}^H\mathbf{N}_1\Delta\mathbf{h}_{1,i}+2Re\lbrace \Delta\mathbf{h}_{1,i}\mathbf{N}_1\bar{\mathbf{h}}_{1,i}^H\rbrace +\bar{\mathbf{h}}_{1,i}^H\mathbf{N}_1\bar{\mathbf{h}}_{1,i}-\\
		(2^\gamma_1-1)\bigg(\sigma_{1,i}^2+\sum_{m=1}^{K_2}\hat{\mathbf{h}}_{2,i}^H\mathbf{V}_m\hat{\mathbf{h}}_{2,i}\bigg) \geq 0.
	\end{aligned}
\end{equation}

By introducing a new variable $\lambda_{1,i}$, we can transform the constraint ($\ref{OP4st1}$) into a linear matrix inequality (LMI) form with Lemma 1, which can be expressed as

\begin{equation}\label{1}
	\begin{aligned}
		{
		\left[ \begin{array}{ccc}
			\mathbf{N}_1+\lambda_{1,i}\mathbf{I}_N & \mathbf{N}_1\bar{\mathbf{h}}_{1,i} \\
			\bar{\mathbf{h}}_{1,i}^H\mathbf{N}_1 & \bar{\mathbf{h}}_{1,i}^H\mathbf{N}_1\bar{\mathbf{h}}_{1,i}-(2^{\gamma_1}-1)c_1-\lambda_{1,i}\epsilon_1^2
		\end{array}
			\right ]}\succeq 0,
	\end{aligned}
\end{equation}
where $c_1=\sum_{m=1}^{K_2}\hat{\mathbf{h}}_{2,i}^H\mathbf{V}_m\hat{\mathbf{h}}_{2,i}+\sigma_{1,i}^2$, and $\lambda_{1,i}$ is the slack variable with $\lambda_{1,i} \geq 0$. Likewise, the constraint ($\ref{OP4st2}$) is transformed into LMI constraint as follows

\begin{equation}\label{2}
	\begin{aligned}
		{
			\left[ \begin{array}{ccc}
				\!\!\mathbf{N}_2+\lambda_{2,m}\mathbf{I}_N &\!\!\!\! \mathbf{N}_2\bar{\mathbf{h}}_{2,m} \\
				\!\!\!\!\!\!\!\!\bar{\mathbf{h}}_{2,m}^H\mathbf{N}_2 &\!\!\!\!\!\!\!\! \bar{\mathbf{h}}_{2,m}^H\mathbf{N}_2\bar{\mathbf{h}}_{2,m}-(2^{\gamma_2}-1)\sigma_{2,m}^2-\lambda_{2,m}\epsilon_2^2
			\end{array}
			\right ]}\succeq 0,
	\end{aligned}
\end{equation}
where $\mathbf{N}_2=\mathbf{V}_m-(2^\gamma_2-1)\sum_{k=1,k\ne m}^{K_2}\mathbf{V}_k$, and $\lambda_{2,m}$ is a new introduced  slack variable with $\lambda_{2,m} \geq 0$. Based on the above transformations, the optimization problem can be rewritten as

\begin{subequations}
	\begin{eqnarray}
		\mathcal{Q}3:	\!\!\!\!\!\!\underset{\mathbf{W}_i,\mathbf{V}_m,\lambda_{1,i},\lambda_{2,m}}{\mathop{\text{min}}}\,\!\!\!\!\!\!\!\!\!\!&&\!\!\!\!\!\sum_{i=1}^{K_1}\alpha_{1,i}\text{tr}(\mathbf{W}_i)+\sum_{m=1}^{K_2}\alpha_{2,m}\text{tr}(\mathbf{V}_m) \label{OP4obj}\\
		\textrm{s.t.}&&\!\!\!\!\!\!\!\!\!\!\! (\ref{1}),\label{OP5st1}\\
		&&\!\!\!\!\!\!\!\!\!\!\! (\ref{2}),\label{OP5st2}\\
		&&\!\!\!\!\!\!\!\!\!\!\! \sum_{i=1}^{K_1}\parallel \mathbf{w}_i\parallel^2\leq P_T,\label{OP5st3}\\
		&&\!\!\!\!\!\!\!\!\!\!\! \bigg[\sum_{m=1}^{K_2}\mathbf{v}_m\mathbf{v}_m^H \bigg]_{k,k}\leq P_S,\quad \forall k \label{OP5st4}\\
		&&\!\!\!\!\!\!\!\!\!\!\! \mathbf{W}_i\succeq 0,\text{rank}(\mathbf{W}_i)=1,\forall i,\label{OP5st5}\\
		&&\!\!\!\!\!\!\!\!\!\!\! \mathbf{V}_m\succeq 0,\text{rank}(\mathbf{V}_m)=1,\forall m\label{OP5st6}
	\end{eqnarray}
\end{subequations}

Problem ($\mathcal{Q}3$) is still nonconvex due to the rank-1 constraints \cite{rank2}. To effectively tackle the rank-1 constraints ($\ref{OP5st5}$) and ($\ref{OP5st6}$) and obtain a high-quality solution for problem ($\mathcal{Q}3$), we propose a penalty-based algorithm. To facilitate the algorithm based on penalty, we need the following lemma.

\begin{figure*}[hb] 
	\centering
	\rule[-10pt]{18cm}{0.07em}
\begin{subequations}
	\begin{eqnarray}
		\mathcal{Q}4:	\!\!\!\!\!\!\underset{\mathbf{W}_i,\mathbf{V}_m,\lambda_{1,i},\lambda_{2,m}}{\mathop{\text{min}}}\,\!\!\!\!\!\!\!\!\!\!&&\!\!\!\!\!\sum_{i=1}^{K_1}\alpha_{1,i}\text{tr}(\mathbf{W}_i)+\sum_{m=1}^{K_2}\alpha_{2,m}\text{tr}(\mathbf{V}_m)+\rho_1 \bigg(\sum_{i=1}^{K_1}\big(\text{tr}(\mathbf{W}_i)-\lambda_{1,i,\max}\big)+\sum_{m=1}^{K_2}\big(\text{tr}(\mathbf{V}_m)-\lambda_{2,m,\max}\big)\bigg) \label{OP6obj}\\
		\textrm{s.t.}&& (\ref{1}),(\ref{2}),(\ref{OP5st3}),(\ref{OP5st4}), \label{OP6st1}\\
		&& \mathbf{W}_i\succeq 0,\forall i,\label{OP6st2}\\
		&& \mathbf{V}_m\succeq 0,\forall m\label{OP6st3}
	\end{eqnarray}
\end{subequations}
\end{figure*}

\emph{Lemma 2:} The rank-1 condition $\text{rank}(\mathbf{M})=1$ for any positive semi-definite $\mathbf{M} \in \mathbb{C}^{N\times N}$ can be expressed as the following condition:

\begin{equation}
	\begin{aligned}
	\text{tr}(\mathbf{M})=\parallel \mathbf{M} \parallel_2,
	\end{aligned}
\end{equation}
where $\parallel \mathbf{M}\parallel_2$ is the spectral norm of $\mathbf{M}$, $\text{tr}(\mathbf{M})=\sum_{i=1}^{N}\lambda_n$, and $\lambda_n$ is the $n$-th largest eigenvalue of matrix $\mathbf{M}$. Note that $\mathbf{W}_i$ and $\mathbf{V}_m$ are all positive semidefinite matrix, which means that each eigenvalue of $\mathbf{W}_i$ and $\mathbf{V}_m$ satisfied $\lambda_n\geq 0$. Intuitively, rank-1 constraints imply that only one eigenvalue is larger than zero. Thus, according to the lemma 2 above, we can replace the rank-1 constraints of ($\ref{OP5st5}$) and ($\ref{OP5st6}$) with the following equation:

\begin{equation}
	\begin{aligned}\label{penalty 1}
		\text{tr}(\mathbf{W}_i)-\sum_{i=1}^{K_1}\lambda_{1,i,\max}=0,
	\end{aligned}
\end{equation}

\begin{equation}
	\begin{aligned}\label{penalty 2}
		\text{tr}(\mathbf{V}_m)-\sum_{m=1}^{K_2}\lambda_{2,m,\max}=0.
	\end{aligned}
\end{equation}
Then, we can rebuild the objective function ($\ref{OP4obj}$) by introducing the penalty function with the constraints ($\ref{penalty 1}$) and ($\ref{penalty 2}$).

\begin{equation}
	\begin{aligned}
	&\underset{\mathbf{W}_i,\mathbf{V}_m,\lambda_{1,i},\lambda_{2,m}}{\mathop{\text{min}}}\,\sum_{i=1}^{K_1}\alpha_{1,i}\text{tr}(\mathbf{W}_i)+\sum_{m=1}^{K_2}\alpha_{2,m}\text{tr}(\mathbf{V}_m)\\&+\rho_1 \bigg(\sum_{i=1}^{K_1}\big(\text{tr}(\mathbf{W}_i)-\lambda_{1,i,\max}\big)+\sum_{m=1}^{K_2}\big(\text{tr}(\mathbf{V}_m)-\lambda_{2,m,\max}\big)\bigg)
	\end{aligned}
\end{equation}
where $\rho_1>0$ denotes the penalty factor used to penalize the violations of the equality constraints ($\ref{penalty 1}$) and ($\ref{penalty 2}$). It is important to note that as $\rho_1$ approaches infinity, the best result of problem ($\mathcal{Q}3$) will always satisfy the rank-1 constraints ($\ref{OP5st5}$) and ($\ref{OP5st6}$), making it the best result of problem ($\mathcal{Q}3$). Therefore, our focus shifts to solving problem ($\mathcal{Q}3$) in the subsequent steps.

Unfortunately, the presence of the penalty function in the new objective function renders it nonconvex. To solve such a nonconvex issue, we adopt a successive convex approximation method to get the convex one. More precisely, for the solutions $\mathbf{W}_i^{(t)}$ and $\mathbf{V}_m^{(t)}$ during the $t$-th iteration, respectively, we can obtain the inequality as below

\begin{equation}
	\begin{aligned}\label{upper bound 1}
	\text{tr}(\mathbf{W}_i^{(t+1)})-(\mathbf{e}_{1,i,\max}^{(t)})^H\mathbf{W}_i^{(t+1)}\mathbf{e}_{1,i,\max}^{(t)} \\ \geq \text{tr}(\mathbf{W}_i^{(t+1)})-\lambda_{1,i,\max}^{(t+1)}\geq 0,
	\end{aligned}
\end{equation}

\begin{equation}
	\begin{aligned}\label{upper bound 2}
		\text{tr}(\mathbf{V}_m^{(t+1)})-(\mathbf{e}_{2,m,\max}^{(t)})^H\mathbf{V}_m^{(t+1)}\mathbf{e}_{2,m,\max}^{(t)} \\ \geq \text{tr}(\mathbf{V}_m^{(t+1)})-\lambda_{2,m,\max}^{(t+1)}\geq 0,
	\end{aligned}
\end{equation}
where $\mathbf{e}_{1,i,\max}$ and $\mathbf{e}_{2,m,\max}$ denote the unit eigenvector corresponding to $\lambda_{1,i,\max}$ and $\lambda_{2,m,\max}$, respectively. Thus, by replacing ($\ref{penalty 1}$) and ($\ref{penalty 2}$) with their upper bound ($\ref{upper bound 1}$) and ($\ref{upper bound 2}$), the optimization problem ($\mathcal{Q}4$) becomes convex, which is shown at the bottom of the page. By using some available optimization software packages, e.g., CVX, it can be effectively solved. The solving process of the problem ($\mathcal{Q}4$) is summarized in Algorithm 1.

\begin{algorithm}[!h]
	\caption{Joint Robust Beamforming for Integrated Satellite-Terrestrial Maritime Communication}
	\label{alg:AOA}
	\renewcommand{\algorithmicrequire}{\textbf{Input:}}
	\renewcommand{\algorithmicensure}{\textbf{Output:}}
	\begin{algorithmic}[1]
		\REQUIRE $K_1$, $M_1$, $K_2$, $M_2$, $P_S$, $P_T$, $\alpha_{1,i}$, $\alpha_{2,m}$, $\gamma_1$, $\gamma_2$, $\sigma^2_0$, $\varepsilon_1$, $\varepsilon_2$  
		\ENSURE $\mathbf{w}_i$, $\mathbf{v}_m$    
		
		\STATE $\mathbf{Initialization:}$ Set positive constant $\tau$, penalty factor $\rho_1$, maximal iteration number $t_{\max}$, accuracy $\epsilon_1$, $\epsilon_2$, feasible solution $\mathbf{W}_i^{(0)}$, $\mathbf{V}_m^{(0)}$ and iteration index $t=0$
		\STATE $\mathbf{repeat}$
		\STATE \ \  Use CVX to deal with problem ($\mathcal{Q}4$), then get $\mathbf{W}_i^{(t)}$ and $\mathbf{V}_m^{(t)}$
		\IF {$|\text{tr}(\mathbf{W}_i^{(t)})-\lambda_{1,i,\max}^{(t)}|>\epsilon_1$ and $|\text{tr}(\mathbf{V}_m^{(t)})-\lambda_{2,m,\max}^{(t)}|>\epsilon_2$}
		\STATE Set $\rho_1=\rho_1+\tau$
		\ENDIF
		\STATE Set $t=t+1$
		\STATE $\mathbf{until}$ $t=t_{\max}$ or problem is solved
		\STATE $\mathbf{Finally}$, apply eigenvalue decomposition (EVD) to $\mathbf{W}_i^{(t)}$, $\mathbf{V}_m^{(t)}$ and get $\mathbf{w}_i$, $\mathbf{v}_m$.
	\end{algorithmic}
\end{algorithm}

\begin{table*}[h!t]
	\small
	\centering
	\caption{Main Simulation System Parameters For LEO Satellite IoT precoding scheme}\label{Simulation}
	\begin{tabular}{|c|c|}
		\hline
		Parts & Complexity in order of ln$\frac{1}{\eta }c_i$  \\\hline
		Part 1 with $n_1=\mathcal{O}(K_1M_1^2)$ & $\hat{c}_1=\sqrt{2K_1(M_1+1)}\cdot n_1\cdot \lbrace K_1[(M_1+1)^3+M_1^3+1]+K_1n_1[(M_1+1)^2+M_1^2+1]+n_1^2$$\rbrace$ \\\hline
		Part 2 with $n_2=\mathcal{O}(K_2M_2^2)$ & $\hat{c}_2=\sqrt{K_2(2M_2+3)}\cdot n_2\cdot \lbrace K_2[(M_2+1)^3+M_2^3+2]+K_2n_2[(M_2+1)^2+M_2^2+2]+n_2^2$$\rbrace$  \\\hline
	\end{tabular}
\end{table*}	

\subsection{Algorithm Analysis}
In this subsection, we analyze the performance of the proposed Algorithm 1 in terms of convergence behavior and computational complexity.

\emph{Convergence Analysis:} For ($\mathcal{Q}4$), each individual optimization variable is convex when the rank-1 constraints are disregarded. As a result, we can obtain the stationary point for each variable and the best solution for the problem during the iterations. By utilizing Algorithm 1, we can ensure that the objective values of problem ($\mathcal{Q}4$) with solution $\mathbf{W}_i^{(t+1)}$ and $\mathbf{V}_m^{(t+1)}$ are always less than or equal to the solution $\mathbf{W}_i^{(t)}$ and $\mathbf{V}_m^{(t)}$. This guarantees the existence of a lower bound for the objective value, indicating that Algorithm 1 will ultimately converge to a minimum value after a certain number of iterations.

\emph{Complexity Analysis:} The problem ($\mathcal{Q}4$) solely consists of linear matrix inequality (LMI) and second-order cone (SOC) constraints \cite{complexity}. Consequently, the computational complexity of the proposed Algorithm 1 can be analyzed by the standard interior-point method (IPM), where the complexity consists of two components: the complexity per iteration and the complexity introduced by iterations. Given a $\eta >0$, the number of iterations needed to achieve a $\eta$-optimal solution to problem ($\mathcal{Q}4$) is approximately proportional to $\zeta \cdot \ln(1/\eta)$, where $\zeta$ represents the barrier parameter to measure the geometric complexity of the conic constraints. During each iteration, a search direction is determined by dealing with a system of $n$ linear equations with $n$ unknowns. Besides, the computation cost is primarily influenced by the construction and factorization of the coefficient matrix of the linear system. We can divide the computational complexity of Algorithm 1 into two  parts which are summarized in Table \uppercase\expandafter{\romannumeral1}. Consequently, the overall complexity of Algorithm 1 is given by $\hat{c}_1+\hat{c}_2$, which can obtain the sub-optimal solutions within polynomial time complexity. The details of the computational complexity expressions are presented in Appendix.

\begin{figure}[t]
	\centering
	\includegraphics[width=0.5\textwidth]{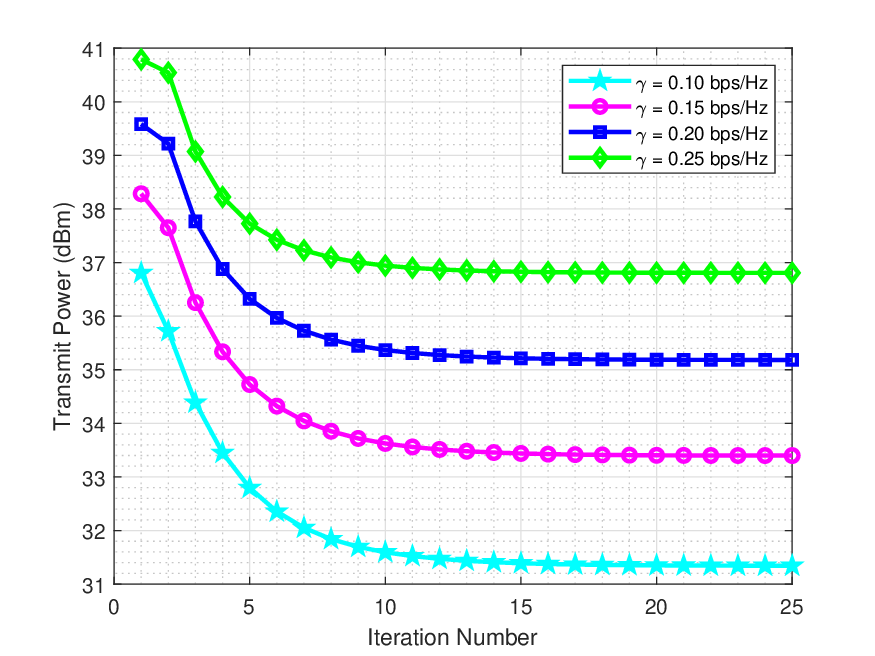} 
	\caption{Convergence performance of the proposed algorithm with different required date rates.}
	\label{diedai}
\end{figure}

\begin{figure}[t]
	\centering
	\includegraphics[width=0.5\textwidth]{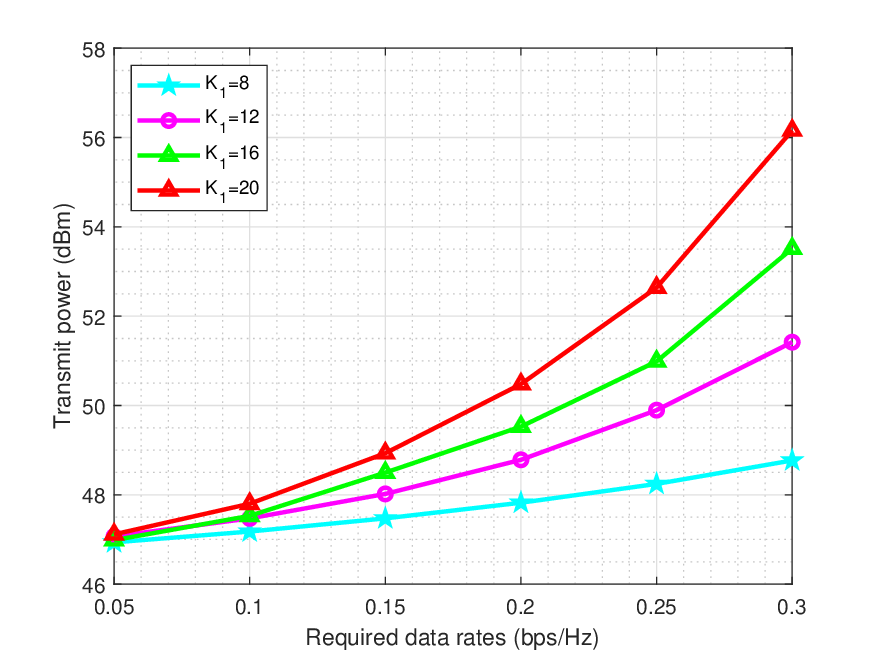} 
	\caption{Total transmit power versus required data rates with different numbers of near-shore users.}
	\label{r_K1}
\end{figure}

\begin{figure}[t]
	\centering
	\includegraphics[width=0.5\textwidth]{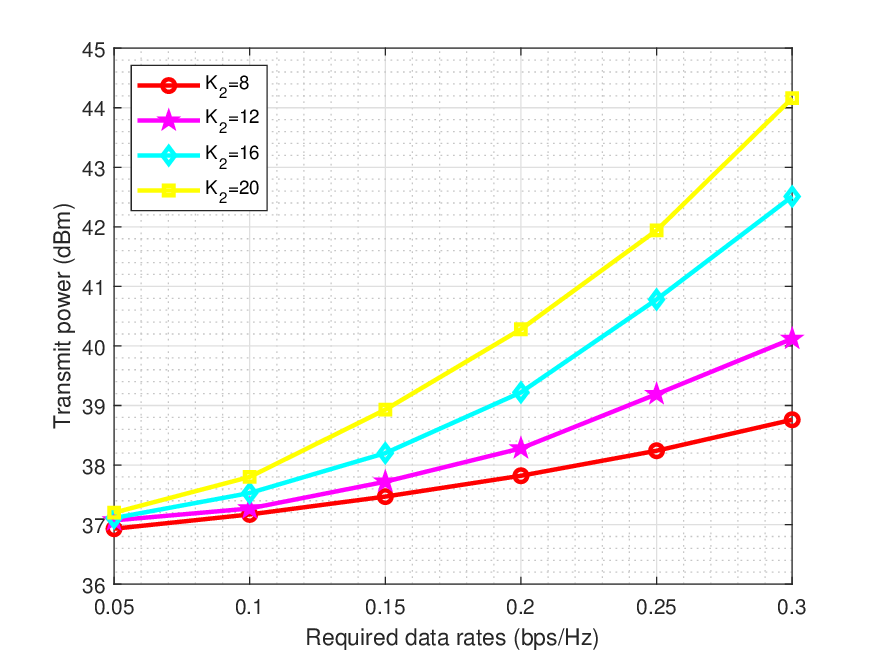} 
	\caption{Total transmit power versus required data rates with different numbers of off-shore users.}
	\label{r_K2}
\end{figure}

\section{Simulation Results}
In this section, we provide extensive simulation results to demonstrate the effectiveness of the proposed robust beamforming design algorithm for integrated satellite-terrestrial maritime communications. The simulation parameters are outlined in Table \uppercase\expandafter{\romannumeral2}.

\begin{table}[t]
	\small
	\centering
	\caption{Main Simulation Parameters For Integrated Satellite-Terrestrial Communication Systems}\label{Simulation}
	\begin{tabular}{|c|c|}
		\hline
		Parameter & Value\\\hline
		Satellite orbit & LEO \\\hline
		Carrier frequency &  160 MHz \\\hline
		Bandwidth  &  20 MHz \\\hline
		Number of TBS antennas & 8 \\\hline
		Number of satellite antennas  & 8\\\hline
		Number of near-shore users  &  4 \\\hline
		Number of off-shore users  &  6 \\\hline
		Satellite antenna gain   &  55 dBi \\\hline
		TBS antenna gain   &  30 dBi \\\hline
		Receive antenna gain  & 20 dBi \\\hline
		Noise temperature   &  300 K \\\hline
		Altitude of orbit      &  550 km  \\\hline
		Variance of AWGN & -110dBm\\ \hline
		Boltzmann's constant  &  1.38$\times 10^{-23}$ J/m\\ \hline
		Rain fading mean &  -2.6dB\\\hline
		Rain fading variance  &  1.63dB \\\hline
		3dB angle & $0.4^{\circ}$ \\\hline
		Rician factor  & 10 \\\hline
		Required data rates of near-shore users & 0.1 bps/Hz \\ \hline
		Required data rates of off-shore users & 0.1 bps/Hz \\  \hline
		TBS total transmit power   & 47dBm \\ \hline
		Effective antennas height of TBS   & 50m \\\hline
		Effective antennas height of users  & 10m \\\hline
	\end{tabular}
\end{table}	

To begin with, we examine the convergence behavior of the proposed Algorithm 1. To support the entertainment and communication needs for maritime passengers, the minimum data rate required is from 1 to 3 Mbps with an ideal data rate of 4 to 8 Mbps \cite{survey}. Therefore, our simulation results setting up 0.1 to 0.3 bps/Hz with the 20 MHz bandwidth can match the practical scenarios. As shown in Fig. $\ref{diedai}$, the total transmit power of the TBS and the LEO satellite gradually decreases and converges to a stable point within 15 iterations, demonstrating the rapid convergence of the proposed Algorithm 1. In addition, with the increment of the data rates required by users, the total transmit power increases accordingly.

\begin{figure}[t]
	\centering
	\includegraphics[width=0.5\textwidth]{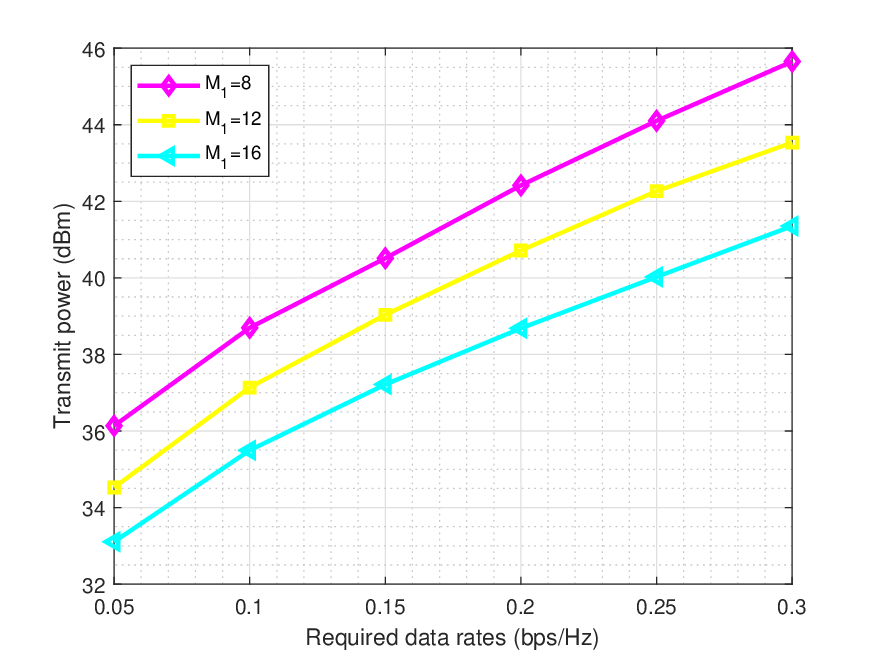} 
	\caption{Total transmit power versus required data rates with different numbers of TBS antennas.}
	\label{r_M1}
\end{figure}

\begin{figure}[t]
	\centering
	\includegraphics[width=0.5\textwidth]{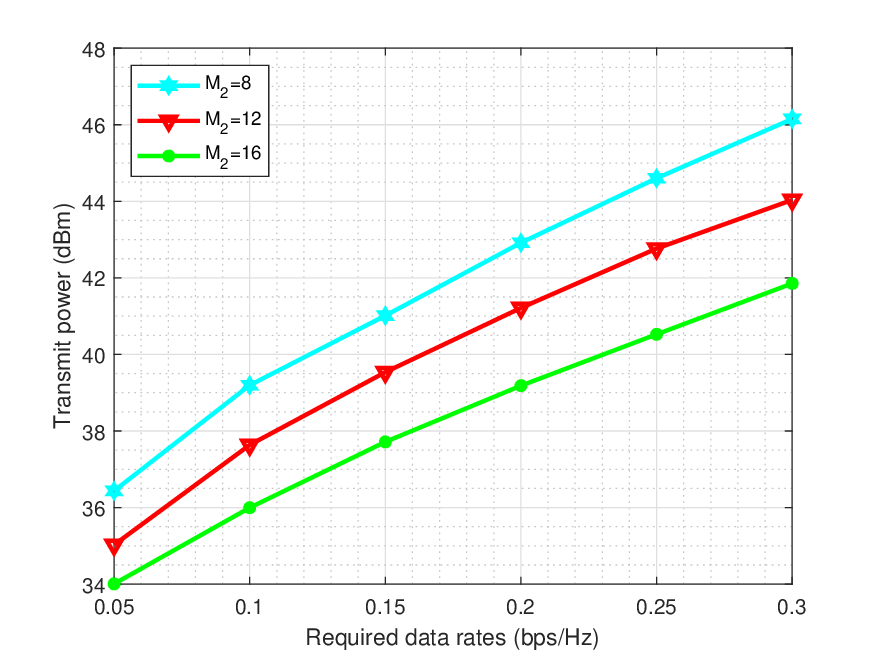} 
	\caption{Total transmit power versus required data rates with different numbers of LEO satellite antennas.}
	\label{r_M2}
\end{figure}

\begin{figure}[t]
	\centering
	\includegraphics[width=0.5\textwidth]{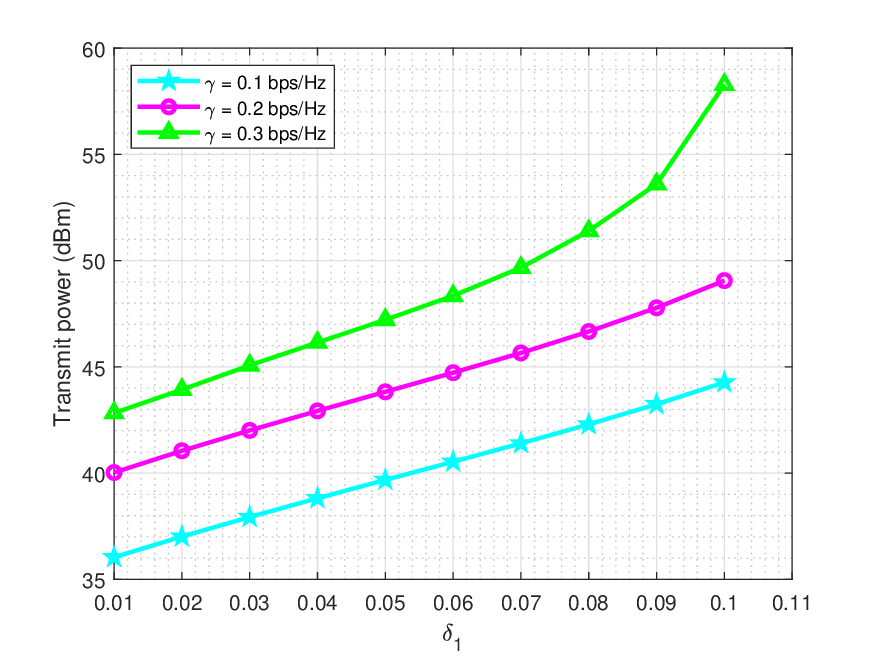} 
	\caption{Total transmit power versus tidal factor $\delta_1$ with different required data rates.}
	\label{wucha1}
\end{figure}

The number of users is also an important factor in the integrated satellite-terrestrial maritime communications. It is shown in Fig. $\ref{r_K1}$ and Fig. $\ref{r_K2}$ that the overall transmit power versus the required data rates with different numbers of near-shore users and off-shore users. It can be seen that the total transmit power rises as the number of users and required data rate increase. This is because when the number of users and the required data rates increase, there may be more interference between users. To satisfy the requirement of data rate, it is necessary to increase the transmit power. In addition, it is found that under the same required data rates, when the number of near-shore users increases, the total transmit power is greater than that of off-shore users. This is because near-shore users not only have interference between neighboring users, but also suffer interference from satellite. Therefore, when the number of users increases, near-shore users need more transmit power to deal with interference.

Furthermore, we investigate the performance of total transmit power on the number of antennas with different required data rates. In Fig. $\ref{r_M1}$ and Fig. $\ref{r_M2}$, it is seen that the total transmit power decreases as the number of antennas increase for all the required data rates, where we set $K_1=K_2=8$. This is because more spatial degrees of freedom can be utilized when the number of antennas increases. Besides, by optimizing the transmit beamforming of the TBS and LEO satellite, it is possible to achieve greater array gains and make more efficient use of the transmit power. In other words, the proposed Algorithm 1 leads to improve efficiency in utilizing the available transmit power.

\begin{figure}[t]
	\centering
	\includegraphics[width=0.5\textwidth]{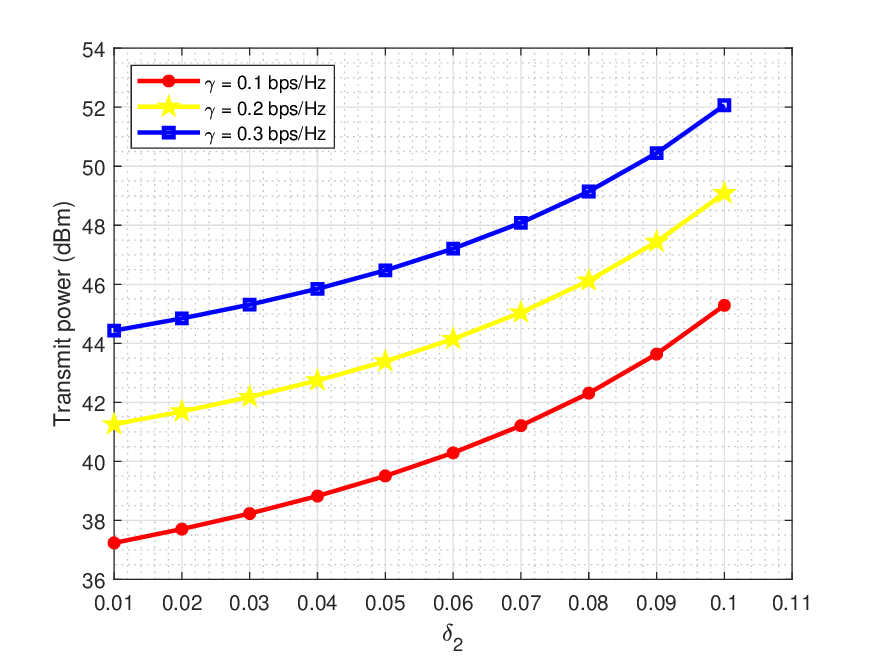} 
	\caption{Total transmit power versus tidal factor $\delta_2$ with different required data rates.}
	\label{wucha2}
\end{figure}

\begin{figure}[t]
	\centering
	\includegraphics[width=0.5\textwidth]{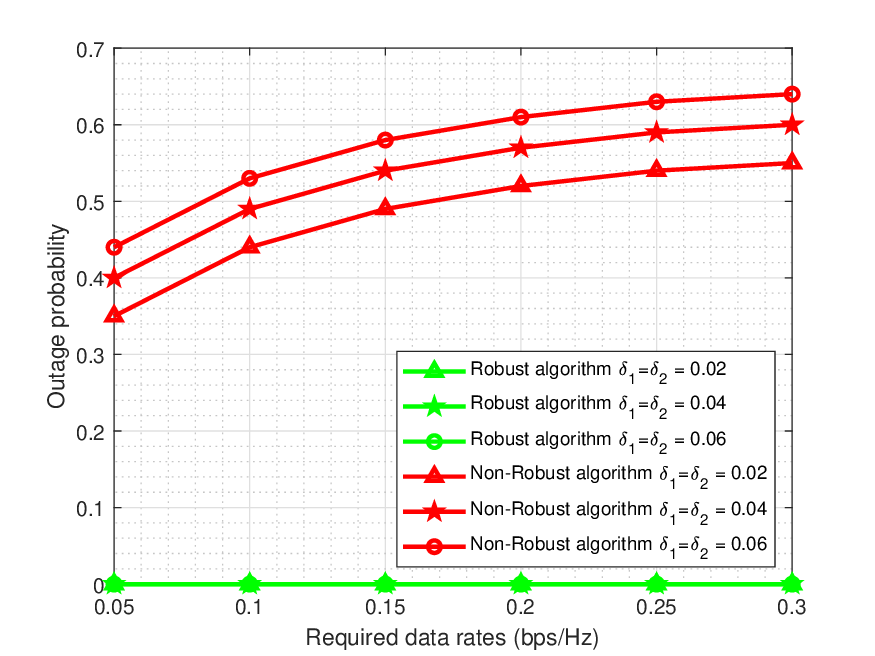} 
	\caption{Outage probabilities of proposed and non-robust algorithms versus required data rates with different values of tidal factor $\delta_1$ and $\delta_2$.}
	\label{outage}
\end{figure}

\begin{figure}[t]
	\centering
	\includegraphics[width=0.5\textwidth]{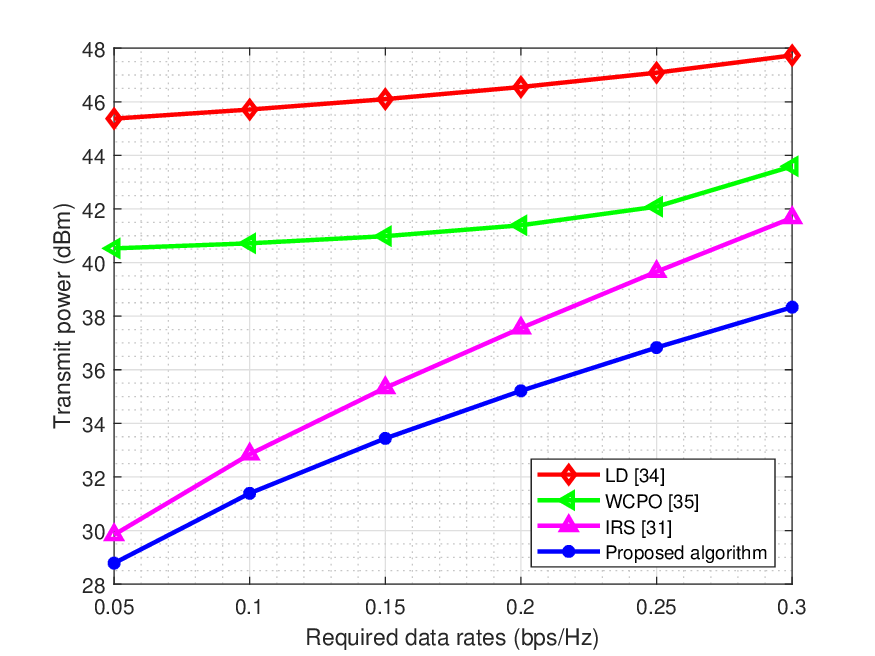} 
	\caption{Comparison of total transmit power with different values of required data rates between proposed algorithm and other algorithms.}
	\label{duibi}
\end{figure}

Fig. $\ref{wucha1}$ and Fig. $\ref{wucha2}$ show the total transmit power versus channel phase error for different required data rate, where the required data rates are set as $r=\gamma_1=\gamma_2$ and numbers of users are set as $K_1=K_2=8$. To describe the CSI error resulting from wave fluctuations, we define a tidal factor related to wave fluctuation severity, which is $\delta_1=\varepsilon_1/\bar{\theta}_1$ and $\delta_2=\varepsilon_2/\bar{\theta}_2$. We can find that the proposed algorithm has high robustness, where the total transmit power performance is good in different tidal factors. When the sea surface environment is extremely harsh, wave fluctuations not only bring about changes in antenna angle, but also cause changes in antenna height. In general, we can well resist the influence of the antenna angle and ignore the influence of the antenna height change. However, when the environment is extremely harsh, that is, when the $\delta_1$ is high, the antenna height change cannot be ignored resulting in the turn-up of simulation curve. Besides, the transmit power exhibits an upward trend as the level of phase uncertainty and required data rates increase, which is also consistent with theoretical analysis.

Next, we compare the outage performance of the non-robust algorithm and proposed algorithm versus the required data rates with different values of tidal factor $\delta_1$ and $\delta_2$. In the case of perfect CSI, the beamforming vector obtained through the non-robust algorithm may fail to meet the minimum data rate requirements of the users due to the CSI errors. The probability of outage is defined as the probability that the minimum data rates required by the users does not meet. The simulation result shows that the outage probability of the proposed algorithm is zero in any case, which proves the effectiveness and robustness of the proposed algorithm.

Finally, we compare the proposed algorithm and some baseline algorithms under the same conditions. The LD algorithm \cite{LD} reformulates the worst-case beamforming problem into a simplified optimization problem by using Lagrange duality. Another WCPO algorithm \cite{WCPO} uses the worst-case performance optimization to solve the power minimization problem. The algorithm of \cite{S} is designed for a small number of near-shore users. When it is applied to the scenario considered in this paper, the number of users and the coverage area increase, thereby more transmit power is required than our proposed algorithm. It can be observed from the Fig. $\ref{duibi}$ that the transmit power of all algorithms increases with the increment of the required data rates, which is also consistent with the previous analysis. Furthermore, the proposed algorithm demonstrates reduced transmit power in comparison to the other algorithms, further substantiating the effectiveness of the proposed algorithm.

\section{Conclusion}
In this paper, we provided an integrated satellite-terrestrial framework for maritime communications with wide coverage. According to the electromagnetic propagation characteristics in maritime communications, we modeled the relationship between actual CSI and obtained CSI. With the obtained CSI, we proposed a joint robust beamforming design algorithm. Both theoretical analysis and simulation results confirmed that the proposed algorithm can minimize the total transmit power compared to other existing algorithms, while guaranteeing the minimum data rate requirements for the maritime users. In addition, simulation results also demonstrate the robustness of the proposed algorithm in the different tidal factors.

\begin{appendices}{
	\section{Derivation of the computational complexity}
The computational complexity of the Algorithm 1 can be divided into two parts, the first part $\hat{c}_1$ can be expressed as
\begin{equation}
	\begin{aligned}
		\hat{c}_1=\sqrt{\beta_1} \cdot (C_1^{\text{form}}+C_1^{\text{fact}}) \cdot \ln(1/\eta),
	\end{aligned}
\end{equation}
where $\beta_1$ is the so-called barrier parameter, which can be computed as

\begin{equation}
	\begin{aligned}
		\beta_1=\underbrace{K_1 \times (M_1+1)}_{\text{due to (\ref{OP5st1})}} + \underbrace{K_1 \times M_1}_{\text{due to (\ref{OP5st5})}}+ \underbrace{K_1 \times 1}_{\text{due to (\ref{OP5st1})}} + \underbrace{1}_{\text{due to (\ref{OP5st3})}},
	\end{aligned}
\end{equation}
where $C_1^{\text{form}}$ is the cost of forming matrix, which can be computed as

\begin{equation}
	\begin{aligned}
		C_1^{\text{form}}=&\underbrace{n_1\cdot K_1 (M_1+1)^3 + n_1^2 \cdot K_1 (M_1+1)^2}_{\text{due to (\ref{OP5st1})}} \\&+ \underbrace{n_1\cdot K_1 M_1^3 + n_1^2 \cdot K_1 M_1^2}_{\text{due to (\ref{OP5st5})}} + \underbrace{n_1\cdot K_1 + n_1^2 \cdot K_1}_{\text{due to (\ref{OP5st3})}},
	\end{aligned}
\end{equation}
where $C_1^{\text{fact}}$ is the cost of factorizing matrix, which can be computed as

\begin{equation}
	\begin{aligned}
		C_1^{\text{fact}}=n_1^3.
	\end{aligned}
\end{equation}

Thus, the first part of complexity $\hat{c}_1$ with $n_1=\mathcal{O}(K_1M_1^2)$ can be computed as
\begin{equation}
	\begin{aligned}
		\hat{c}_1=&\sqrt{2K_1(M_1+1)}\cdot n_1\cdot \lbrace K_1[(M_1+1)^3+M_1^3+1]\\&+K_1n_1[(M_1+1)^2+M_1^2+1]+n_1^2\rbrace,
	\end{aligned}
\end{equation}
similarly, the second part of complexity $\hat{c}_2$ with $n_2=\mathcal{O}(K_2M_2^2)$ can be computed as
\begin{equation}
	\begin{aligned}
		\hat{c}_2=&\sqrt{K_2(2M_2+3)}\cdot n_2\cdot \lbrace K_2[(M_2+1)^3+M_2^3+2]\\&+K_2n_2[(M_2+1)^2+M_2^2+2]+n_2^2\rbrace.
	\end{aligned}
\end{equation}
}
\end{appendices}

\end{document}